\documentclass[multphys,vecphys]{svmult}

\usepackage{makeidx}     
\usepackage{graphicx}    
\usepackage{multicol}    

\makeindex             

\begin{document}
\newcommand{\beq}{\begin{equation}}\newcommand{\oper}[4]{\mbox{$O_{#1}^{#2}(#3,#4)$}}
\newcommand{\eeq}{\end{equation}}
\newcommand{\intif}{\int_{-\infty}^{\infty}}
\newcommand{\sid}{\mbox{$\psi^{\dagger}$}}
\newcommand{\sib}{\mbox{$\overline{\psi}$}}
\newcommand{\il}{\int_{-\Lambda}^{\Lambda}}
\newcommand{\ie}{\int_{0}^{\Lambda}}
\newcommand{\iT}{\int_{0}^{2\pi}d\theta}
\newcommand{\iK}{\int_{\Lambda /K_F}^{\pi -\Lambda /K_F}}
\newcommand{\si}[2]{\mbox{$\psi_{#1}(#2)$}}
\newcommand{\etab}{\mbox{\boldmath $\eta $}}
\newcommand{\sigmab}{\mbox{\boldmath $\sigma $}}
\newcommand{\w}{{\omega}}
\def\half{{1\over 2}}
\def\a{{\alpha}}
\def\b{{\beta}}
\def\g{{\gamma}}
\def\d{{\delta}}
\def\t{{\theta}}
\def\tu{{\tilde u}}
\def\cE{{\cal E}}
\def\cP{{\cal P}}
\def\cT{{\cal T}}
\def\cL{{\cal L}}
\def\cN{{\cal N}}
\def\ve{{\varepsilon}}
\def\e{{\varepsilon}}
\def\ua{\uparrow}
\def\da{\downarrow}
\def\a{{\alpha}}
\def\b{{\beta}}
\def\g{{\gamma}}
\def\d{{\delta}}
\def\de{{\varepsilon}}
\def\t{{\theta}}
\def\bK{{\mathbf K}}
\def\bm{{\mathbf m}}
\def\bsig{{\mathbf \sigma}}
\def\bB{{\mathbf B}}
\def\bp{{\mathbf p}}
\def\bI{{\mathbf I}}
\def\bn{{\mathbf n}}
\def\bM{{\mathbf M}}
\def\bq{{\mathbf q}}
\def\br{{\mathbf r}}
\def\bs{{\mathbf s}}
\def\bS{{\mathbf S}}
\def\bQ{{\mathbf Q}}
\def\bs{{\mathbf s}}
\def\bB{{\mathbf B}}
\def\bl{{\mathbf l}}
\def\bPi{{\mathbf \Pi}}
\def\bJ{{\mathbf J}}
\def\bR{{\mathbf R}}
\def\bz{{\mathbf z}}
\def\ba{{\mathbf a}}
\def\bk{{\mathbf k}}
\def\bP{{\mathbf P}}
\def\bg{{\mathbf g}}
\def\bX{{\mathbf X}}
\def\prl{Phys. Rev. Lett.}
\def\prb{Phys. Rev. {\bf B}}
\def\prd{Phys. Rev.{\bf D}}
\def\pre{Phys. Rev.{\bf E}}
\def\rmp{Rev. Mod. Phys.}
\def\a{{\alpha}}
\def\b{{\beta}}
\title*{The renormalization group approach - from Fermi liquids to quantum dots}
 \titlerunning{RG for interacting fermions}
\author{R.Shankar\inst{1}}
\institute{Sloane Physics Lab\\ Yale University\\ New Haven CT
06520 \\
\texttt{r.shankar@yale.edu} }
%
%
\maketitle

\section{The RG: what, why and how}
\label{sec:1}
%
 Imagine that you have
some problem in the form of a partition function

\beq Z(a,b) = \int dx \int dy e^{-a(x^2 + y^2)} e^{-b (x+y)^4}
\eeq where $a,b$ are the parameters.

First consider $b=0$, {\em the gaussian model}. Suppose  that you
are just interested in $x$, say in its fluctuations.  Then you
have the option of integrating out $y$ and working with the new
partition function \beq Z(a) =N\ \int dx  e^{-a x^2 }\eeq where
$N$ comes from doing the $y$-integration. We will ignore such an
$x$-independent pre-factor here and elsewhere since it will cancel
in any averaging process.

Consider now the nongaussian case with $b\ne 0$. Here we have

\begin{eqnarray}
Z(a', b' ...) &= &\int dx \left[ \int dy e^{-a(x^2 + y^2)} e^{-b (x+y)^4} \right] \nonumber \\
& \equiv       &     \int dx e^{-a'\ x^2 } e^{- b'x^4 - c' x^6 +
...}
\end{eqnarray}
 where $a'$, $b'$ etc., define the parameters of the effective field theory  for $x$.
   These parameters will reproduce exactly the same averages for $x$ as the original  ones.
    {\em This evolution of parameters with the elimination of uninteresting degrees of freedom,
    is what we mean these days by renormalization, and as such has nothing to do with infinities;
    you just saw it happen in a problem with just two variables.}

    The parameters $b$, $c$ etc., are called {\em couplings} and the
    monomials they multiply are called {\em interactions}. The
    $x^2$ term is called the {\em kinetic} or {\em free-field} term.

Notice that to get the effective theory we need to do a
nongaussian integral. This can only be done perturbatively. At the
simplest {\em tree Level}, we simply drop $y$ and find $b'=b$. At
higher orders, we bring down the nonquadratic exponential and
integrate in $y$ term by term and generate effective interactions
for $x$. This procedure can be represented by Feynman graphs in
which variables in the loop are limited  to the ones being
eliminated.

Why do we do this? Because certain tendencies of $x$ are not so
apparent when $y$ is around, but surface to the top, as we zero in
on $x$. For example, we are going to consider a problem in which
$x$ stands for low-energy variables and $y$ for high energy
variables. Upon integrating out high energy variables a
numerically small coupling can grow in size (or initially
impressive one diminish into oblivion), as we zoom in on the low
energy sector.

This notion can be made more precise as follows. Consider the
gaussian model in which we have just $a\ne 0$. We have seen that
this value does not change as $y$ is eliminated since $x$ and $y$
do not talk to each other. This is called a {\em fixed point of
the RG}. Now turn on new couplings  or "interactions"
(corresponding to higher powers of $x$, $y$ etc.) with
coefficients $b$, $c$ and so on. Let $a'$, $b'$ etc., be the new
couplings after $y$ is eliminated. The mere fact that $b'>b$ does
not mean $b$ is more important for the physics of $x$. This is
because $a'$ could also be bigger than $a$.  So we rescale $x$ so
that the kinetic  part, $x^2$,  has the same coefficient as
before. If the quartic term still has a bigger coefficient, (still
called $b'$), we say it is a {\em relevant} interaction. If $b'<b$
we say it is irrelevant. This is because in reality $y$ stands for
many variables, and as they  are eliminated one by one, the
coefficient of the quartic term will run to zero. If a coupling
neither grows not shrinks it is called {\em marginal}.

There is another excellent reason for using the RG, and that is to
understand the phenomenon of universality in critical phenomena. I
must regretfully pass up the  opportunity to explain this and
refer you to Professor Michael Fisher's excellent lecture notes in
this very same school many years ago \cite{MEF}.

We will now see how this method is applied to interacting fermions
in $d=2$. Later we will apply these methods to quantum dots.

\section{The problem of interacting fermions}

Consider a system of nonrelativistic spinless fermions in two
space dimensions. The one particle hamiltonian is \beq H = {K^2
\over 2m} - \mu \eeq where the chemical potential $\mu$  is
introduced to make sure we have a finite density of particles in
the ground state: all levels  up the Fermi surface, a circle
defined by \beq
K^{2}_{F} /2m = \mu \eeq
 are now occupied and occupying these levels lowers the ground-state energy.

Notice that this system has gapless excitations above the ground
state. You can take an electron just below the Fermi surface and
move it just above, and this costs as little energy as you please.
Such a system will carry a dc current in response to a dc voltage.
An important  question one  asks is if this will be true when
interactions are turned on. For example the system could develop a
gap and become an insulator. What really happens for the $d=2$
electron gas?

\begin{figure}
\centering
\includegraphics[height=4cm]{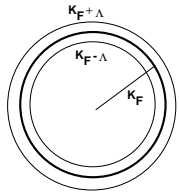}
%
%
\caption{The low energy region for nonrelativistic fermions lies within the annulus concentric with the Fermi circle.}
\label{annulus}       
\end{figure}

We are going to answer this  using the RG. Let us first learn how
to do RG for noninteracting fermions.  To understand the low
energy physics, we take a band of of width $\Lambda$  on either
side of the   Fermi surface. This is the first great difference
between this problem and the usual ones in relativistic field
theory and statistical mechanics. Whereas in the latter examples
low energy means small momentum, here it means small deviations
from the Fermi surface. Whereas in these  older problems we zero
in on the origin in momentum space, here we zero in on a surface.
The low energy region is shown in Figure \ref{annulus}.

To apply our methods we need to  cast the problem  in the form of
a  path integral. Following any number of sources, say
\cite{rmprg} we obtain the following expression for the partition
function of free fermions:

\beq
Z_0= \int  d\psi d\overline{\psi} e^{S_0}\\
\eeq where \beq S_0= \int d^2K \intif d\omega
\overline{\psi}(\omega , \bK) \left(i\omega -
{(K^2-K_{F}^{2})\over 2m}\right)\psi (\omega , \bK)\label{Z0} \eeq
where $\psi$ and $ \overline{\psi}$ are called Grassmann
variables. They are really weird objects one gets to love after
some familiarity. Most readers can assume they are ordinary
integration variables. The dedicated reader can learn more from
Ref. \cite{rmprg}.

We now adapt this general expression to  the annulus to obtain
\beq
Z_0= \int  d\psi d\overline{\psi} e^{S_0}\\
\eeq where \beq S_0= \int_{0}^{ 2\pi}d\theta \intif d\omega \il dk
\overline{\psi} (i\omega - v\ k)\psi \label{Z1}. \eeq  To get here
we have had to approximate as follows: \beq {K^2-K_{F}^{2}\over
2m}\simeq {K_F\over m}\cdot k=v_F\  k\eeq where $k-K-K_F$ and $v_F$ is the fermi
velocity, hereafter set equal to unity. Thus $\Lambda$ can be
viewed as a momentum or energy cut-off measured from the Fermi
circle. We have also replaced $KdK$ by $K_F dk$ and absorbed $K_F$
in $\psi$ and $\overline{\psi}$. It will seen that neglecting $k$
in relation to $K_F$ is irrelevant in the technical sense.

Let us now perform mode elimination and reduce the cut-off by a
factor $s$. Since this is a gaussian integral, mode elimination
just leads to a multiplicative constant we are not interested in.
So the result is just the same action as above, but with $|k| \le
\Lambda /s$. Let us now do make the following additional
transformations:

\begin{eqnarray}
(\omega ', k')                          &=& s(\omega , k)
\nonumber \\
(\psi ' (\omega ', k'), \overline{\psi}' (\omega ', k')) &=&
s^{-3/2} (\psi ({\omega '\over s}, {k'\over s})
 , \overline{\psi} ({\omega '\over s}, {k'\over s})).
\label{rescale}
\end{eqnarray}

When we do this, the action and the phase space all return to
their old values. So what? Recall that our plan is to evaluate the
role of  quartic interactions in low energy physics as we do mode
elimination.    Now what really matters is not the absolute size
of the quartic term, but its size relative to the quadratic term.
Keeping   the quadratic term identical before and after the RG
action   makes the comparison easy: if the quartic coupling grows,
it is relevant; if it decreases, it is irrelevant, and if it stays
the same it is marginal. The system is clearly gapless if the
quartic coupling is irrelevant. Even a marginal coupling  implies
no gap since  any gap will grow under the various rescalings of
the RG.

Let us now turn on a generic four-Fermi interaction in
path-integral form: \beq S_4 = \int \overline{\psi}(4)
\overline{\psi}(3) \psi (2)\psi(1) u(4,3,2,1)\label{s4} \eeq where
$\int $ is a shorthand: \beq \int \equiv \prod_{i=1}^{3} \int{d
\theta_{i}}\int_{- \Lambda}^{\Lambda} dk_{i} \intif d\omega_{i}
\eeq

At the tree level, we simply keep the modes within the new
cut-off, rescale fields, frequencies  and momenta , and read off
the new coupling. We find

\beq u'(k',\omega' , \theta ) = u(k'/s, \omega' /s, \theta)
\label{tree} \eeq

This is the evolution of the coupling function. To deal with
coupling  constants with which we are more familiar, we expand the
functions  in a Taylor series (schematic)

\beq u = u_o + k  u_1  + k^2  u_2 ...
\end{equation}
where $k$ stands for all the $k$'s  and $\omega$'s. An expansion
of this kind is possible since
 couplings in the Lagrangian are nonsingular in a problem with short range interactions.
If we now make such an expansion and compare coefficients in Eqn.
(\ref{tree}),  we find that  $u_0$ is marginal and the rest are
irrelevant, as is any coupling of more than four fields. Now this
is exactly what happens in  $\phi^{4}_{4}$, scalar field theory in
four dimensions with a quartic interaction. The difference here is
that we still have dependence on the angles on the Fermi surface:
$$u_0 = u(\theta_1 , \theta_2 , \theta _3 , \theta_4 ) $$

Therefore in this theory we are going to get coupling functions
and not a few coupling constants.

\begin{figure}
\centering
\includegraphics[height=8cm]{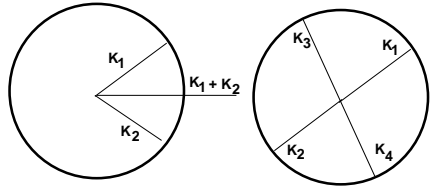}
%
%
\caption{Kinematical reasons why momenta are either conserved
pairwise or restricted to the BCS channel. }
\label{kinematics}       
\end{figure}

Let us analyze this function. Momentum conservation should allow
us to eliminate one angle. Actually it allows us more because of
the fact that these momenta do not come form the entire plane, but
a  very thin annulus near $K_F$. Look at the left half of Figure
\ref{kinematics}. Assuming that the cutoff has been reduced to the
thickness of the circle in the figure, it is clear that if two
points  $1$ and $2$ are chosen from it to represent the incoming
lines in a four point coupling, the outgoing ones are forced to be
equal to them (not in their sum, but individually) up to a
permutation, which is irrelevant for spinless fermions. Thus we
have in the end just one function of two angles, and by rotational
invariance, their  difference:

\beq u(\theta_1 , \theta_2 , \theta _1 , \theta_2 ) = F(\theta_1 -
\theta_2 )  \equiv F(\theta ). \eeq About forty years ago Landau
came to the very same conclusion\cite{landau} that a Fermi system
at low energies would be described by one function defined on the
Fermi surface. He did this without the benefit of the RG and for
that reason, some of the leaps were hard to understand.  Later
detailed diagrammatic calculations justified this picture
\cite{agd}. The RG provides yet another way to understand it. It
also tells us other things,  as we will now see.

The first thing is that the final angles  are not slaved to the
initial ones if the former are exactly  opposite, as in the right
half of Figure 2. In this case, the final ones can be anything, as
long as they are opposite to each other. This leads to one more
set of marginal couplings in the BCS channel, called

\beq u(\theta_1 , -\theta_1 , \theta _3 , -\theta_3 ) = V(\theta_3
- \theta_1) \equiv  V(\theta ). \eeq

The next point is that since $F$ and $V$ are marginal at tree
level, we have to go to one loop to see if they are still so. So
we draw the usual diagrams shown in Figure 3. We eliminate an
infinitesimal momentum slice of thickness $d\Lambda$ at $k= \pm
\Lambda$.

\begin{figure}
\centering
\includegraphics[height=8cm]{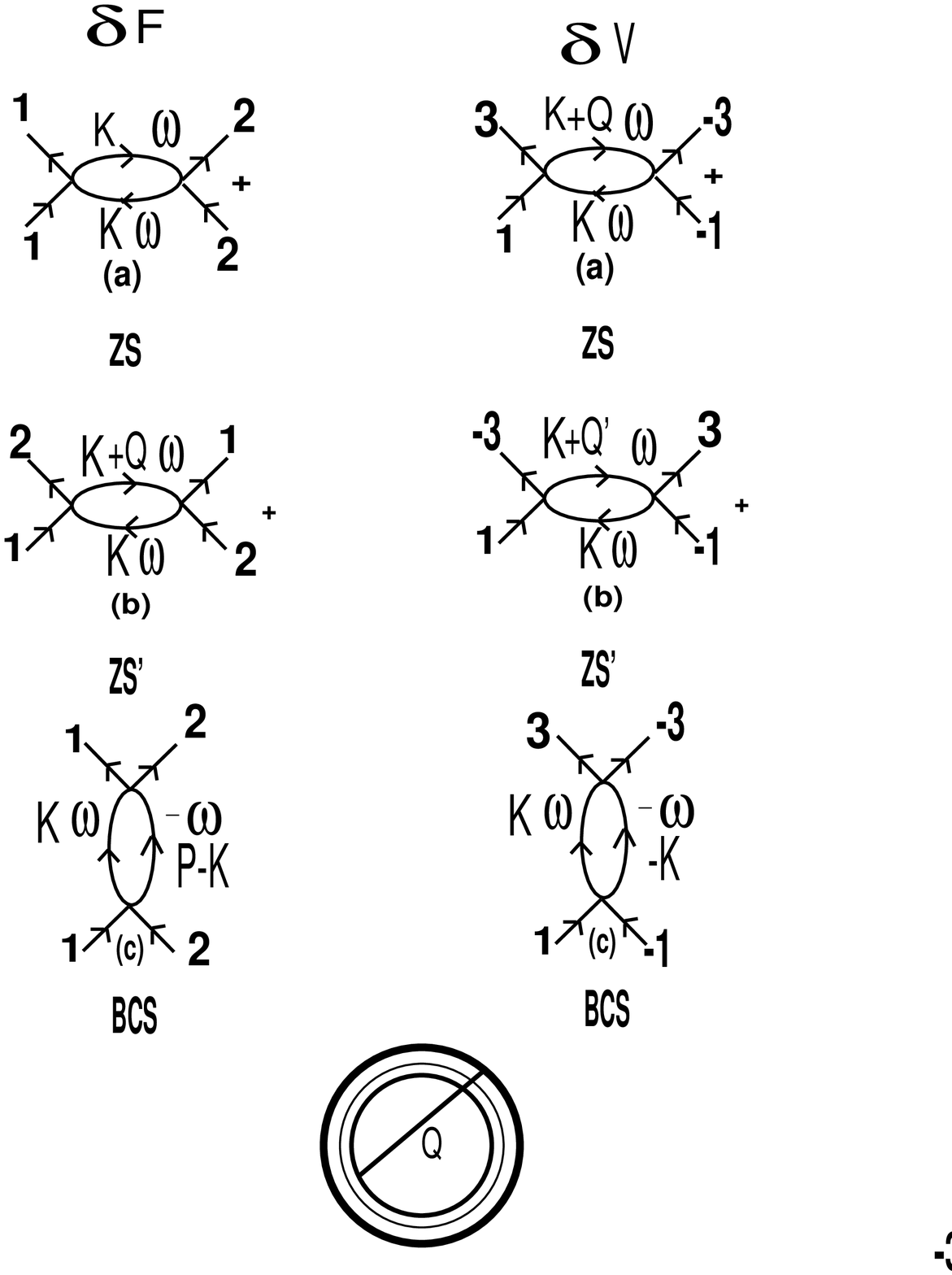}
%
%
\caption{One loop diagrams for the flow of $F$ and $V$. The last at the bottom shows that a
large momentum $Q$ can be absorbed only at two particular initial angles (only one of which is shown)
  if the final state is to lie in the shell being eliminated.  }
\label{rgdiags}       
\end{figure}

These diagrams are like the ones in any quartic field theory, but
each one behaves differently from the others and its its
traditional counterparts. Consider the first one (called ZS) for
$F$. The external momenta have zero frequencies and lie of the
Fermi surface since $\omega$ and $k$ are irrelevant. The momentum
transfer is exactly zero. So the integrand has the following
schematic form:

\beq \delta F \ \simeq \int d{\theta} \int dk d\omega \left(
\frac{1}{(i\omega - \varepsilon(K))}\ \frac{1}{(i\omega  -
 \varepsilon(K))} \right)
 \label{denom}
 \eeq
The loop momentum $K$ lies in one of the two shells being
eliminated.  Since there is no energy difference between the two
propagators, the poles in $\omega$ lie in the same half-plane and
we get zero, upon closing the contour in the other half-plane. In
other words, this diagram can contribute if it is a particle-hole
diagram, but given zero momentum transfer we cannot  convert a
hole at $-\Lambda$ to a particle at $+\Lambda$. In the ZS'
diagram, we have a large momentum transfer, called $Q$ in the
inset at the bottom. This is of order $K_F$ and much bigger than
the radial cut-off, a phenomenon unheard of in say $\phi^4$
theory, where all momenta and transfers are bounded by  $\Lambda$.
This in turn means that the loop momentum is not only restricted
in the direction to a sliver $d\Lambda$, but also in the angular
direction in order to be able to absorb this huge momentum $Q$ and
land up in the other shell being eliminated (see bottom of Fig.
(\ref{rgdiags}). So we have $du \simeq dt^2$, where $dt = d\Lambda
/\Lambda$. The same goes for the BCS diagram. Thus $F$ does not
flow at one loop.

Let us now turn to the renormalization of $V$. The first two
diagrams are useless for the same reasons as before, but the last
one is special. Since the total incoming momentum is zero, the
loop  momenta are equal and opposite and no matter what direction
$K$ has, $-K$ is guaranteed to lie in the same shell being eliminated. However the
loop frequencies are now equal and opposite so that the poles in
the two propagators now lie in opposite half-planes. We now get a
flow (dropping constants)

\beq {dv(\theta_{1} - \theta_{3} ) \over dt} = -  \int d\theta
v(\theta_{1}- \theta )\ v(\theta - \theta_{3} ) \label{bcs} \eeq

Here is an example of a flow equation for a coupling function.
However by expanding in terms of angular momentum eigenfunctions
we get an  infinite number of flow equations

\beq \frac{dv_m}{dt} = - v_{m}^2.\label{bcsrg} \eeq one for each
coefficient. These equations tell us  that if the potential in
angular momentum channel $m$ is repulsive, it will get
renormalized down to zero ( a result derived many years ago by
Anderson and Morel) while if it is attractive, it will run off,
causing the BCS instability.  This is the reason the $V$'s are not
a part of Landau theory, which assumes we have no phase
transitions. This is also a nice illustration of what was stated
earlier: one could begin with a large positive coupling, say $
v_3$ and a tiny negative coupling $v_5$. After much
renormalization, $v_3$ would  shrink to a tiny value and $v_5$
would dominate.
\section{Large-$N$ approach to Fermi liquids}

Not only did Landau say we could describe Fermi liquids with an
$F$ function, he also managed to compute the  response  functions
at small $\omega$ and $q$  in terms of the $F$ function even when
it was large, say $10$, in dimensionless units. Again the  RG
gives us one way to understand this. To this end we need to recall
the the key ideas of "large-N" theories.

These theories  involve interactions between $N$ species of
objects. The largeness of $N$ renders fluctuations (thermal or
quantum) small, and enables one to make approximations which are
not perturbative in the coupling constant, but are controlled by
the additional small parameter $1/N$.

As a specific example  let us consider the Gross-Neveu
model\cite{gross-neveu} which is one of the simplest fermionic
large-$N$ theories. This theory has $N$ identical massless
relativistic fermions interacting through a  short-range
interaction. The Lagrangian density is

\beq {\cal{L}}=\sum\limits_{i=1}^{N}{\bar \psi}_i\not\!\partial
\psi_i -{\lambda\over N}
\bigg(\sum\limits_{i=1}^{N}{\bar\psi}_i\psi_i\bigg)^2
\label{gnl}\eeq

Note that the kinetic term  conserves the internal index, as does
the interaction term: any index that goes in comes out. You do not
have to know much about the GN model to to follow this discussion,
which is all about the internal indices.

\begin{figure}
\centering
\includegraphics[height=6cm]{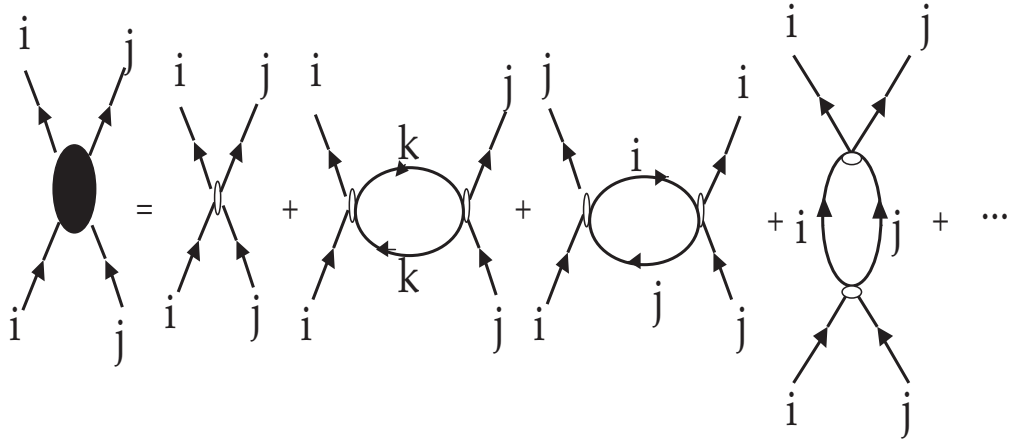}
%
%
\caption{Some diagrams from a large-N theory}
\label{oneoverN}       
\end{figure}

Figure \ref{oneoverN} shows the first few diagrams in the
expression for the scattering amplitude of particle of isospin
index  $i$ and $j$  in the Gross-Neveu theory. The ``bare'' vertex
comes with a factor $\lambda/N$. The one-loop diagrams all share a
factor $\lambda^2/N^2$ from the two vertices. The first one-loop
diagram has a free internal summation over the index $k$ that runs
over $N$ values, with the contribution being identical for each
value of $k$. Thus, this one-loop diagram acquires a compensating
factor of $N$ which makes its contribution of order $\lambda^2/N$,
the {\it same order in $1/N$ as the bare vertex}. However, the
other one-loop diagrams have no such free internal summation and
their contribution is indeed of order $1/N^2$. Therefore, to
leading order in $1/N$, one should keep only diagrams which have a
free internal summation for every vertex, that is, iterates of the
leading one-loop diagram, which are called bubble graphs. {\em For
later use remember that in the diagrams that survive (do not
survive), the indices $i$ and $j$ of the incoming particles do not
(do) enter the loops.}  Let us assume that the momentum integral
up to the cutoff $\Lambda$ for one bubble gives a factor
$-\Pi(\Lambda, q_{ext})$, where $q_{ext}$ is the external momentum
or frequency transfer at which the scattering amplitude is
evaluated. To leading order in large-$N$ the full expression for
the scattering amplitude is

\beq \Gamma (q_{ext})={1\over N}{\lambda\over1+\lambda
\Pi(\Lambda, q_{ext})} \eeq

Once one has the full expression for the scattering amplitude (to
leading order in $1/N$) one can ask for the RG flow of the
``bare'' vertex as the cutoff is reduced by demanding that the
physical scattering amplitude $\Gamma$ remain insensitive to the
cutoff. One then finds (with $t=\ln(\Lambda_0/\Lambda)$)

\beq {d\Gamma_{}(q_{ext})\over dt}=0\Rightarrow {d\lambda\over
dt}=-\lambda^2 {d\Pi(\Lambda, q_{ext})\over dt} \eeq which is
exactly the flow one would extract at one loop. Thus   the
one-loop RG flow is the exact answer to leading order in a
large-$N$ theory. All higher-order corrections must therefore be
subleading in $1/N$.

\subsection{Large-N applied to Fermi liquids}

Consider now  the ${\bar \psi} \psi - {\bar \psi} \psi$ correlation function (with
vanishing values of external frequency and momentum transfer).
Landau showed that it  takes the form \beq \chi = {\chi_0\over
1+F_0},\eeq where $F_0$ is the angular average of $F(\theta )$ and
$\chi_0$ is the answer when $F=0$. Note that the answer is not
perturbative in $F$.

Landau got this result by working with  the ground-state energy as
a functional of Fermi surface deformations. The RG provides us
with not just the ground-state energy, but an effective
hamiltonian (operator)  for all of low-energy physics. This
operator problem can be solved using large $N$-techniques.

The value of $N$ here is of order $K_F/\Lambda$, and here is how
it enters the formalism. Imagine dividing the annulus in Fig.
({\ref{annulus}) into $N$ patches of size $(\Lambda)$ in the
radial and angular directions. The momentum of each fermion ${\bf
k}_i$ is a sum of a "large" part ($\mathcal{O}(k_F))$ centered on
a patch labelled by a patch index $i=1,...N$ and a "small"
momentum $(\mathcal{O}(\Lambda) $ within the patch\cite{rmprg}.

Consider a four-fermion Green's function, as in Figure
(\ref{oneoverN}). The incoming momenta are labelled by the patch
index (such as $i$) and the small momentum is not shown but
implicit. We have seen that as $\Lambda \to 0$, the two outgoing
momenta are equal to the two incoming momenta up to a permutation.
At small but finite $\Lambda$ this means the patch labels are same
before and after. Thus the patch index plays the role of a
conserved isospin index as in the Gross-Neveu model.

The electron-electron interaction terms, written in this notation,
(with $\bk$ integrals  replaced by a sum over patch index and
integration over small momenta) also come with a pre-factor of
$1/N$ ($\simeq \Lambda /K_F$).

It can then be verified that in all Feynman diagrams of this
cut-off theory the patch index plays the role of the  conserved
isospin index exactly as in a theory with $N$ fermionic species.
For example in Figure (\ref{oneoverN}) in the first diagram, the
external indices  $i$ and $j$ do not enter the diagram (small
momentum transfer only) and so the loop momentum is nearly same in
both lines and integrated fully over the annulus, i.e., the patch
index $k$ runs over all $N$ values.  In the second diagram, the
external label $i$ enters the loop and there is a large momentum
transfer ($\mathcal{O}({K_F}))$. In order for both momenta in the
loop to be within the annulus, and to differ by this large $q$,
the angle of the loop momentum is limited to a range
$\mathcal{O}(\Lambda /K_F)$. (This just means that if one momentum
is from patch $i$ the other has to be from patch $j$. ) Similarly,
in the last loop diagram, the angle of the loop momenta is
restricted to one patch. In other words, the requirement that all
loop momenta in this cut-off theory lie in the annulus singles out
only diagrams that survive in the large $N$ limit.

 The
sum of bubble diagrams, singled out by the usual large-$N$
considerations, reproduces Landau's Fermi liquid theory. For
example in the case of $\chi$, one obtains a geometric series that
sums to give $\chi = {\chi_0\over 1+F_0}.$

Since  in the large $N$ limit, the one-loop $\beta$-function for
the fermion-fermion coupling is exact, it follows that the
marginal nature of the Landau parameters $F$ and the flow of $V$ ,
Eqn. (\ref{bcsrg}), are both exact as $\Lambda \to 0$.

A long  paper of mine  Ref. (\cite{rmprg}) explains all this,   as
well as how it is to be generalized to anisotropic Fermi surfaces
and Fermi surfaces with additional special features and
consequently additional instabilities.  Polchinski \cite{pol}
independently analyzed the isotropic Fermi liquid (though not in
the same detail, since it was a just paradigm or toy model for an
effective field theory for him).

\section{Quantum dots}
We will now apply some of these ideas, very successful in the
bulk,  to   two-dimensional quantum dots\cite{qd-reviews,aleiner}- tiny
spatial regions of size $L\simeq 100-200 nm$, to which electrons
are restricted using gates. The dot can be connected weakly or
strongly to leads. Since many experts on dots are contributing to
this volume, I will be sparing in   details and
 references.

Let us get acquainted with some energy scales, starting with
$\Delta$, the mean single particle level spacing. The Thouless
energy is defined as $E_T= {\hbar / \tau}$, where $\tau$ is the
time it takes to traverse the dot. If the dot is strongly coupled
to leads, this is the uncertainty in the energy of an electron as
it traverses the dot. Consequently the $g$ (sharply defined)
states of an isolated dot within $E_T$ will contribute to
conductance and lead to a (dimensionless) conductance $g={E_T\over
\Delta}$.

The dots in question have two features important to us. First,
motion within the dot is ballistic: $L_{el}$, the elastic
scattering length is the same as $L$, the dot size, so that
$E_T={\hbar v_F\over L}$, where $v_F$ is the Fermi velocity. Next,
the boundary of the dot is sufficiently irregular as to cause
chaotic motion at the classical level. At the quantum level
single-particle energy levels and wavefunctions (in any basis)
within $E_T$ of the Fermi energy will resemble those of a random
hamiltonian matrix and be described Random Matrix Theory (RMT)
\cite{RMT}. We will only invoke a few  results from RMT and they
will be explained in due course.

At a generic value of gate voltage $V_g$ the ground state has a
definite number of particles $N$ and energy $\mathcal{E}_N$. If
${\mathcal{E}}_{N+1}-{\mathcal{E}}_N=\alpha eV_g$ ($\alpha$ is a
geometry-dependent factor) the  energies of the $N$ and
$N+1$-particle states are degenerate, and a tunnelling peak occurs
at zero bias. Successive peaks are separated by the second
difference of ${\mathcal{E}}_N$, called $\Delta_2$, the
distribution of which is measured. Also measured are statistics of
peak-height
distributions\cite{peak-height-th},\cite{peak-height-expt,peak-height-expt2},
which depend on wavefunction statistics of RMT.

To describe the data one needs to write down a suitable
hamiltonian \beq H_U=\sum\limits_{\a } \e_\a c^{\dagger}_{\a }
c_{\a}+{1\over 2}\sum_{\a \b \g \d}V_{\a \b \g
\d}c^{\dag}_{\a}c^{\dag}_{\b}c^{}_{\g}c^{}_{\d}\eeq (where the
subscripts label the exact single particle states including spin)
and try to extract its implications.  Earlier theoretical
investigations were confined to the noninteracting limit: $V\equiv
0$ and missed  the fact that due to the small capacitance of the
dot, adding an electron required some significant charging energy
on top of the energy of order $\e_\a$ it takes to promote an
electron by one level. Thus efforts have been made to include
interactions\cite{coulomb-blockade,H_U-prehistory,H_U-kurland,aleiner,qd-numerics}.

The simplest model  includes a constant charging energy $U_0N^2/2$
\cite{coulomb-blockade,qd-reviews}. Conventionally $U_0$ is
subtracted away  in plotting $\Delta_2$. This model predicts a
bimodal distribution for $\Delta_2$: Adding an electron above a
doubly-filled (spin-degenerate) level costs $U_0+\e$, with $\e$
being the energy to the next single-particle level. Adding it to a
singly occupied level costs $U_0$. While the second contribution
gives a  delta-function peak at $0$ after $U_0$ has been
subtracted, the first contribution is the distribution of nearest
neighbor level separation $\e$,  of the order of  $\Delta$. But
simulations \cite{qd-numerics} and
experiments\cite{small-rs-expt,large-rs-expt} produce
distributions for $\Delta_2$ which do not show any bimodality, and
are much broader.

The next significant advance was the discovery of the Universal
Hamiltonian \cite{H_U-prehistory,H_U-kurland}. Here one keeps only
couplings of the form $V_{\a \b \b \a}$ on the grounds that  only
they have a non-zero ensemble average (over disorder
realizations). This seems reasonable in the limit of large $g$
since  couplings with zero average are typically of size $1/g$
according to RMT. The Universal Hamiltonian is thus \beq
H_U=\sum\limits_{\a,s} \e_\a c^{\dagger}_{\a,s} c_{\a,s}
+{U_0\over2}{N}^2-{J\over2}{\bS}^2 +\lambda\bigg(\sum\limits_{\a}
c^{\dagger}_{\a,\ua}c^{\dagger}_{\a,\da}\bigg)\bigg(\sum\limits_{\b}c_{\b,\da}c_{\b,\ua}\bigg)
\label{hu}\eeq  where  $s$ is single-particle spin and ${\bS }$ is
the total spin. The Cooper coupling $\lambda$ does not play a
major role, but the inclusion of the exchange coupling $J$ brings
the theoretical
predictions\cite{H_U-prehistory,H_U-kurland,aleiner} into better
accord with experiments, especially if one-body
``scrambling''\cite{scrambling,adametal,2brim,tau2} and finite
temperature effects are taken into account. However, some
discrepancies still remain in  relation to
numerical\cite{qd-numerics} and experimental
results\cite{large-rs-expt}  at  $r_s\ge2$.

We now see that the following dot-related  questions naturally
arise. Given that  adding more  refined interactions (culminating
in the universal hamiltonian) led to better descriptions of the
dot, should one not seek  a more systematic way to to decide what
interactions should be included from the outset? Does our past
experience with clean systems and bulk systems tell us how to
proceed? Once we have written down a comprehensive hamiltonian, is
there a way to go beyond perturbation theory to unearth
nonperturbative physics in the dot, including possible phases and
transitions between them? What will be the experimental signatures
of these novel phases and the transitions between them if indeed
they do exist? These questions will now be addressed.

\subsection{Interactions and Disorder: Exact results on the dot}

The first crucial   step towards this goal was taken by  Murthy
and Mathur \cite{qd-us1}. Their ideas was as follows.
\begin{itemize}
\item  {\bf Step 1:} {\em Use the clean system RG  described earlier
\cite{rmprg} (eliminating momentum states on either side of the
Fermi surface ) to eliminate all states far from the Fermi surface
till one  comes down to the Thouless band, that is, within $E_T$
of $E_F$. }

We have seen that this process inevitably leads to Landau's Fermi
liquid interaction (spin has been suppressed): \beq V=
\sum\limits_{m=0}^{\infty}{u_m\Delta\over2}\sum\limits_{k,k'}
\cos[m(\t-\t')] c^{\dagger}_{\bk}c_{\bk}c^{\dagger}_{\bk'}c_{\bk'}
\label{landau} \eeq where $\t,\t'$ are the angles of $\bk,\ \bk'$
on the Fermi circle, and $u_m$ is defined by \beq F(\theta ) = \sum_m
u_m e^{im\theta }. \eeq

A few words before we proceed.  First, some experts will point out
that the interaction one gets from the RG allows for small
momentum transfer, i.e., there should be an additional sum over a
small values  $q$ in Eqn. (\ref{landau}) allowing $\bk \to \bk +\bq$ and $\bk'\to \bk'-\bq$. It can be shown that in
the large $g$ limit this sum has just one term, at $q=0$. Unlike
in a clean system, there is no singular behavior associated with
$q\to 0$ and this assumption is a good one. Others  have asked how one can introduce the Landau
interaction that respects momentum conservation in a dot that does
not conserve momentum or anything else except energy. To them I
say this. Just think of a pair of molecules colliding in a room.
As long as the collisions take place in a time scale smaller than
the time between collisions with the walls, the interaction will
be momentum conserving. That this is true for a collision in the
dot for particles moving at $v_F$, subject an interaction of range
equal to the Thomas Fermi screening length (the typical range) is
readily demonstrated. Like it or not, momentum is a special
variable even in a chaotic but ballistic dot since it is tied to
translation invariance, and that that is operative for realistic
collisions within the dot.

\item  {\bf Step 2:}  {\em Switch to the exact basis states of the
chaotic dot, writing the kinetic and interaction terms in this
basis. Run the RG by eliminating exact energy eigenstates within
$E_T$. }

\end{itemize}

While this looks like a reasonable plan, it is not clear how it is
going to be executed since knowledge of the exact eigenfunctions is needed
to even write down the   Landau interaction in the disordered
basis:
\begin{eqnarray}
V_{\alpha \beta \gamma \delta}=& {\Delta\over 4}\sum\limits_{\bk
\bk'}u (\theta -\theta' ) \left[ \phi^{*}_{\alpha}(\bk )
\phi^{*}_{\beta}(\bk' )
-\phi^{*}_{\alpha}(\bk' ) \phi^{*}_{\beta}(\bk )\right] \nonumber\\
&\times \left[ \phi_{\gamma}(\bk' ) \phi_{\delta}(\bk )
-\phi_{\gamma}(\bk ) \phi_{\delta}(\bk' )\right]
\label{V-disorder-basis}
 \end{eqnarray}
where $\bk$ and $\bk'$ take $g$ possible values. These are chosen
as follows. Consider the momentum states of energy within $E_T$ of
$E_F$. In a dot momentum is defined with an uncertainty $\Delta k
\simeq 1/L$ in either direction. Thus one must form packets in $k$
space obeying this condition. It can be easily shown that $g$ of
them will fit into this band. One way to pick such packets is to
simply take plane waves of precise $\bk$ and chop them off at the
edges of the dot and normalize the remains. The $g$ values of
$\bk$  can be chosen with an angular spacing $2\pi /g$. It can be
readily verified that such states are very nearly orthogonal. The
wavefunction  $\phi_{\delta}(\bk )$ is the projection of exact dot
eigenstate $\delta$ on the state $\bk$ as defined above.

We will see that one can go a long way without detailed knowledge
of the wavefunctions $\phi_{\delta}(\bk )$.

First, one can take  the view of the Universal Hamiltonian (UH)
adherents and consider the ensemble average (enclosed in $<>$ ) of
the interactions.
 RMT tells us that to leading order in $1/g$,

\begin{eqnarray}
<\phi_\a^*(\bk_1)\phi_\b^*(\bk_2)\phi_\g(\bk_3)\phi_\d(\bk_4)>&=
{\d_{\a\d}\d_{\b\g}\d_{\bk_1\bk_4}\d_{\bk_2\bk_3}\over g^2}
+{\d_{\a\g}\d_{\b\d}\d_{\bk_1\bk_3}\d_{\bk_2\bk_4}\over
g^2}\nonumber
\\ &+{\d_{\a\b}\d_{\g\d}\d_{\bk_1,-\bk_2}\d_{\bk_3,-\bk_4}\over g^2}
\label{4pt1} \end{eqnarray}

It is seen that only matrix elements in Eqn.
(\ref{V-disorder-basis}) for which the indices $\a\b\g\d$ are
pairwise equal survive disorder-averaging, and also that the
average has no dependence on the energy of $\a\b\g\d$. In the
spinless case, the first two terms on the right hand side make
equal contributions and produce the constant charging energy in
the Universal Hamiltonian of Eq. (\ref{hu}), while in the spinful
case they produce the charging and exchange terms. The final term
of Eq. (\ref{4pt1}) produces the Cooper interaction of Eq.
(\ref{hu}).

{\em Thus the UH contains the rotationally invariant part of the
Landau interaction.} The others, i.e., those that do not survive
ensemble averaging,  are dropped because they  are of order $1/g$.
But we have seen before in the BCS instability of the Fermi liquid
that a term that is nominally small to begin with can grow under
the RG. That this is what happens in this case was shown by the RG
calculation of  Murthy and Mathur. There was however one catch.
The neglected couplings could overturn the UH description for
couplings that exceeded   a critical value. However the critical
value
  is of order unity and so one could not trust either the location or
  even the very existence of this critical point based on their
  perturbative one-loop calculation.
  Their work also gave no clue as to what lay on the other side of
  the critical point.

  Subsequently Murthy and I \cite{qd-us2} showed
  that the methods of the large $N$ theories (with $g$
  playing the role of $N$) were applicable here and could be used
  to show nonperturbatively in the interaction strength that the
  phase transition indeed exists.  This approach  also allowed
 us to  study in detail the phase on the other side of the
 transition, as well as what is called the quantum critical region, to be described later.

\begin{figure}
\centering
\includegraphics[height=4cm]{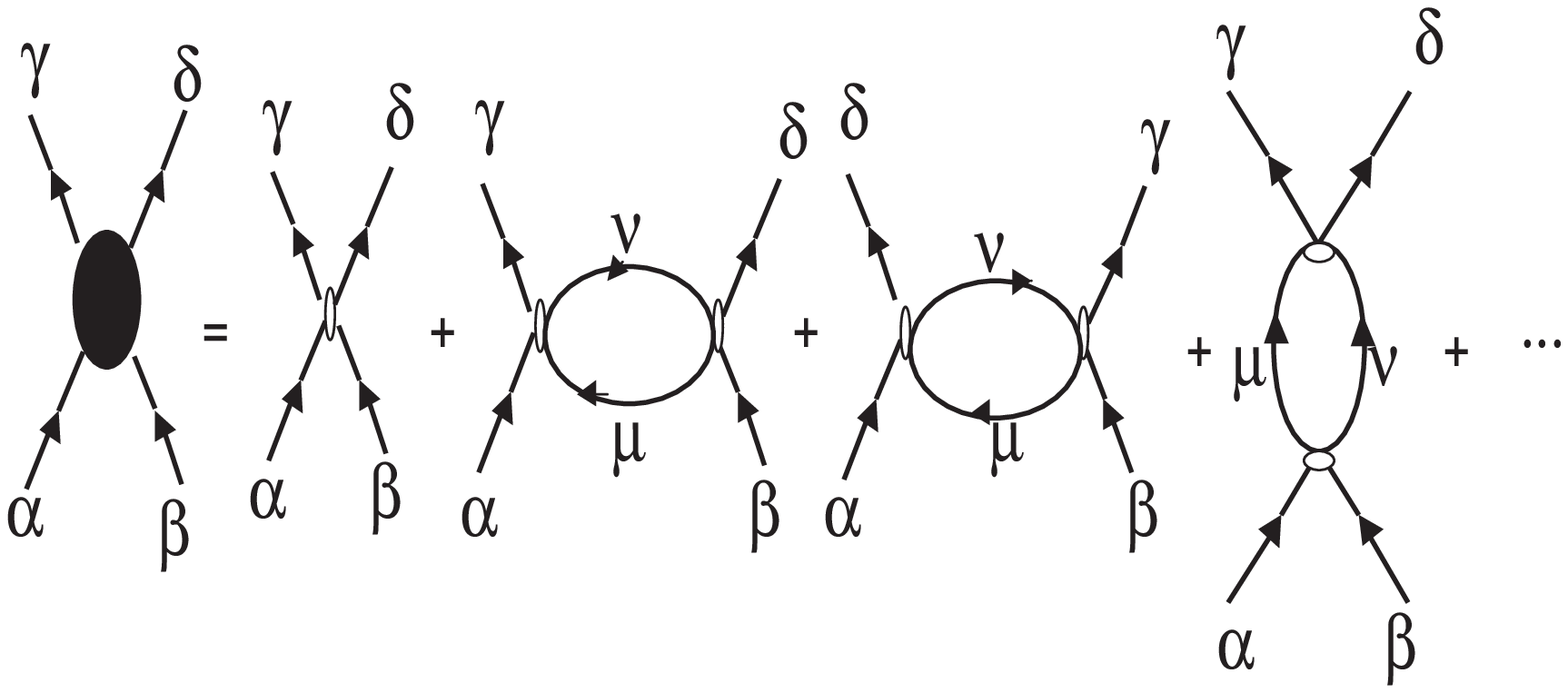}
%
%
\caption{Feynman diagrams for the full four-point amplitude
$\Gamma_{\a\b\g\d}$.}
\label{scatt-amplitude1}       
\end{figure}

Let us now return to
  Murthy and Mathur and ask how the RG flow is derived.
   After integrating some of the $g$ states within $E_T$, we end up with
$g'=ge^{- t }$ states. Suppose  we compute a scattering amplitude
$\Gamma_{\a\b\g\d}$ for the process in which two fermions
originally in states $\a\b$ are scattered into states $\g\d$. This
scattering can proceed directly through the vertex $V_{\a\b\g\d}(
t )$, or via intermediate virtual states higher order in the
interactions, which can be classified by a set of Feynman
diagrams, as shown in Figure \ref{scatt-amplitude1}. All the
states in the diagrams belong to the $g'$ states kept.  {\em  We
demand that the entire amplitude be independent of $g'$, meaning
that the physical amplitudes should be the same in the effective
theory as in the original theory. } This will lead to a set of
flow equations for the $V_{\a\b\g\d}$. In principle this flow
equation will involve all powers of $V$ but we will keep only
quadratic terms (the one-loop approximation). Then the diagrams
are limited to the ones shown in Figure \ref{scatt-amplitude1},
leading to the following contributions to the scattering amplitude
$\Gamma_{\a\b\g\d}$

\begin{eqnarray} \Gamma_{\a\b\g\d}=&V_{\a\b\g\d}\nonumber \\
+&\sum\limits_{\mu,\nu}^{} {}'
{N_F(\nu)-N_F(\mu)\over\ve_{\mu}-\ve_{\nu}}
(V_{\a\nu\mu\d}V_{\b\mu\nu\g}-V_{\a\nu\mu\g}V_{\b\mu\nu\d})\nonumber
\\ -&\sum\limits_{\mu\nu}^{}
{}'{1-N_F(\mu)-N_F(\nu)\over\ve_{\mu}+\ve_{\nu}}
V_{\a\b\mu\nu}V_{\nu\mu\g\d} \label{rgflow} \end{eqnarray} where
the prime on the sum reminds us that only the $g'$ remaining
states are to be kept and where $N_F(\a)$ is the Fermi occupation
of the state $\a$. We will confine ourselves to zero temperature
where this number can only be zero or one. The matrix element
$V_{\a\b\g\d}$ now explicitly depends on the RG flow parameter $ t
$.

Now we demand that upon integrating the two states at $\pm
g'\Delta/2$ we recover the same $\Gamma_{\a\b\g\d}$. Clearly,
since $g'=ge^{- t }$,

\beq {d\over d t }=-g'{\d\over\d g'} \label{xisign}\eeq

The effect of this differentiation on the loop diagrams is to fix
one of the internal lines of the loop to be at the cutoff $\pm
g'\Delta/2$, while the other one ranges over all smaller values of
energy. In the particle-hole diagram, since $\mu$ or $\nu$ can be
at $+g'\Delta/2$ or $-g'\Delta/2$, and the resulting summations
are the same in all four cases, we take a single contribution and
multiply by a factor of 4. The same reasoning applies to the
Cooper diagram. Let us define the energy cutoff
$\Lambda=g'\Delta/2$ to make the notation simpler. Since we are
integrating out two states we have $\d g'=2$

\begin{eqnarray} 0=&{dV_{\a\b\g\d}\over d t }\nonumber \\
-&{g'\over 2}4\sum\limits_{\mu=\Lambda,\nu}^{} {}'
{N_F(\nu)-N_F(\mu)\over\ve_{\mu}-\ve_{\nu}}
(V_{\a\nu\mu\d}V_{\b\mu\nu\g}-V_{\a\nu\mu\g}V_{\b\mu\nu\d})\nonumber
\\ +&{g'\over 2}4\sum\limits_{\mu=\Lambda,\nu}^{}
{}'{1-N_F(\mu)-N_F(\nu)\over\ve_{\mu}+\ve_{\nu}}
V_{\a\b\mu\nu}V_{\nu\mu\g\d} \label{rgflow2} \end{eqnarray} where
$\mu =\Lambda$ means $\varepsilon_\mu =\Lambda$ and so on. The
changed sign in front of the 1-loop diagrams reflects the sign of
Eq. (\ref{xisign})

So far we have not made any assumptions about the form of
$V_{\a\b\g\d}$, and the formulation applies to any finite system.
In a generic system such as an atom, the matrix elements depend
very strongly on the state being integrated over, and the flow
must be followed numerically for each different set $\a\b\g\d$
kept in the low-energy subspace.

In our problem things have become so bad that are good once again:
the wavefunctions $\phi (\bk )$ that enter the matrix elements above
have so scrambled up by disorder that
they can be handled by RMT. In particular it is possible to argue
that the sum over the $g'$ terms may be replaced by it ensemble
average. In other words the flow equation is self-averaging. While
the most convincing way to show this is to compute its variance,
and see that it is of order $1/\sqrt{g}$ times its average, this
fact can be motivated in the following way: There is a sum over
$g'\gg1$ values of $\nu$ with a slowly varying energy denominator,
which makes the sum over $\nu$ similar to a spectral average,
which in RMT is the same as an average over the disorder ensemble.
A more sophisticated argument is presented in Ref \cite{gang4}. We
can use the result
\begin{eqnarray}
<\phi^*_\mu(\bp_1)\phi^*_\nu(\bp_2)\phi_\nu(\bp_3)\phi_\mu(\bp_4)>=\nonumber\\
{\d_{14}\d_{23}\over g^2}-{\d_{13}\d_{24}\over
g^3}-{\d_{1,-2}\d_{3,-4}\over g^3} \label{crucial1}\end{eqnarray}
to deal with the product of four wavefunctions inside the loop and
deal with the energy sum as follows: \beq
\sum\limits_{\ve_\nu=-\Lambda}^{0}{1\over\Lambda+|\ve_\nu|}\approx
\int\limits_{0}^{\Lambda}{d\ve\over\Delta}{1\over
\Lambda+\ve}={\ln{2}\over\Delta} \eeq We are exploiting the fact
that wavefunction correlations are energy independent in the large
-$g$ limit.

After we make this simplifications we find that there are many
kinds of terms of which one kind dominates in the large -$g$
limit.

Let us go back to the properly antisymmetrized matrix element
defined in terms of the Fermi liquid interaction function, Eq.
(\ref{V-disorder-basis}). Since there is a product of two $V$'s in
each loop diagram, and each $V$ contains 4 terms, it is clear that
each loop contribution has 16 terms.  Let us first consider a term
of the leading type in the particle-hole diagram, which survives
in the large-$g$ limit. Putting in the full wavefunction
dependencies (and ignoring factors other than $g,\ g'$) we have
the following contribution from this type of term

\begin{eqnarray} {dV_{\a\b\g\d}\over d t }_{\mbox{Leading}} = &
g'\Delta^2\sum\limits_{\nu=-\Lambda}^{0} {1\over\Lambda+|\ve_\nu|}
\sum\limits_{\bk\bk'}\sum\limits_{\bp\bp'}
u(\t_\bk-\t_\bp)u(\t_{\bp'}-\t_{\bk'})\nonumber \\
& \times \phi^*_\a(\bk)\phi^*_\b(\bk')\phi_\g(\bk')\phi_\d(\bk)
\phi^*_\mu(\bp)\phi^*_\nu(\bp')\phi_\nu(\bp)\phi_\mu(\bp')
\label{diagram-leading}\end{eqnarray}


 Substituting the appropriate  momentum labels for
the particle-hole diagram in Eqn. (\ref{crucial1}), we see that the wavefunction average
relevant to the sum over intermediate states is
\beq {\d_{\bp\bp'}\over g^2}-{1+\d_{\bp,-\bp'}\over g^3}
\label{crucial2}\eeq

Using the self-averaging, the first term of Eq.(\ref{crucial2})
forces $\bp=\bp'$ in Eq. (\ref{diagram-leading}). For large $g$,
using \beq \sum\limits_{\bp} =g\int {d\t_\bp\over 2\pi} \eeq we
obtain  a convolution of the two Fermi liquid functions \beq
\sum\limits_{\bp}u(\t_\bk-\t_\bp)u(\t_\bp-\t_{\bk'})=
g\big(u_0^2+\half\sum\limits_{m=1}^{\infty}
u_m^2\cos{m(\t-\t')}\big) \label{convolve}\eeq where we have
reverted to the notation $\t=\t_\bk,\ \t'=\t_{\bk'}$. In the
second term of Eq. (\ref{crucial2}), the $\d_{\bp,-\bp'}$ turns
out to be subleading, while the other allows independent sums over
$\bp,\ \bp'$. This means that only $u_0$ contributes to this term, (other avrerage to zero
upon summing over all angles)
which produces

\beq
-\sum\limits_{\bp\bp'}u(\t_\bk-\t_\bp)u(\t_{\bp'}-\t_{\bk'})=g^2
u_0^2 \label{convolve-u0}\eeq

  Feeding this into full expression for this contribution
to the particle-hole diagram, we find it to be \begin{eqnarray}
{dV_{\a\b\g\d}\over d t }_{\mbox{Leading}}=&{g'\over g}\Delta
\ln{2}\sum\limits_{\bk\bk'}\bigg(\sum\limits_{m=1}^{\infty}
u_m^2\cos{m(\t-\t')}\bigg)\nonumber\\
&\phi^*_\a(\bk)\phi^*_\b(\bk')\phi_\g(\bk')\phi_\d(\bk)
\end{eqnarray}
 Notice that the result is still of the
Fermi liquid form. In other words  the couplings
$V_{\a\b\g\d}$ which were written in terms of Landau parameters
$u_m$, flow into renormalized coupling once again expressible in  terms of renormalized Landau
parameters. By comparing the two sides, we see each $u_m$ flows
independently of the others as per \beq {du_m\over
dt}=-e^{-t}(\ln{2})u_m^2\ \ \ \ \ \ \ \ m\ne 0
 \label{rg-spinless1}\eeq

  The above
equation can be written in a more physically transparent form by
using a rescaled  variable (for $m\ne0$ only)

\beq \tu_m=e^{-t} u_m \eeq

in terms of which the flow equation becomes

\beq {d\tu_m\over dt}=-\tu_m-(\ln{2})\tu_m^2\equiv \beta(\tu_m)
\label{rg-spinless2}\eeq where the last is a definition of the
$\b$-function.

The reason  $u_o$ does not flow is that the
corresponding interaction
  commutes with the one-body ``kinetic'' part, and
therefore does not suffer quantum fluctuations.

This is the answer at large $g$. We have dropped subleading
contributions of the following type:

\begin{eqnarray}{dV_{\a\b\g\d}\over d t }|_{\mbox{sub}}  &=-g'\Delta^2 \sum\limits_{\nu=-\Lambda}^{0}
{1\over\Lambda+|\ve_\nu|} \left[
\sum\limits_{\bk\bk'}\sum\limits_{\bp\bp'}
u(\t_\bk-\t_\bp)u(\t_{\bp'}-\t_{\bk'})\ \right. \nonumber \\
&\left. \times
\phi^*_\a(\bp)\phi^*_\b(\bk')\phi_\g(\bk')\phi_\d(\bk)
\phi^*_\mu(\bk)\phi^*_\nu(\bp')\phi_\nu(\bp)\phi_\mu(\bp')\right]
\label{diagram-subleading}\end{eqnarray}

Note that the momentum labels of $\phi^*_\a$ and $\phi^*_\mu$ have
been exchanged compared to Eq. (\ref{diagram-leading}) and there
is a minus sign, both coming from the antisymmetrization of Eq.
(\ref{V-disorder-basis}). Once again we ensemble average the
internal $\mu,\ \nu$ sum, the wavefunction part of which gives

\begin{equation}
<\phi^*_\mu(\bk)\phi^*_\nu(\bp')\phi_\nu(\bp)\phi_\mu(\bp')>=
{\d_{\bk\bp'}\d_{\bp\bp'}\over
g^2}-{\d_{\bk\bp}+\d_{\bk,-\bp'}\d_{\bp,-\bp'}\over g^3}
\end{equation}

It is clear that there is an extra momentum restriction in each
term compared to Eq. (\ref{crucial2}), which means that one can no
longer sum freely over $\bp$ to get the factor of $g$ in Eq.
(\ref{convolve}), or the factor of $g^2$ in Eq.
(\ref{convolve-u0}). Therefore this contribution will be down by
$1/g$ compared to that of Eq. (\ref{diagram-leading}).

Turning now to the Cooper diagrams, the internal lines are once
again forced to have the same momentum labels as the external
lines by the Fermi liquid vertex, therefore they do not make any
leading contributions.

 {\em The general rule is that whenever a
momentum label corresponding to an internal line in the diagram
(here $\mu$ and $\nu$) is forced to become equal to a momentum
label corresponding to an external disorder label (here $\a , \b ,
\g , $ or $\d$), the diagram is down by $1/g$, exactly as in the
$1/N$ expansion.} The fact that $1/g$ plays the role of $1/N$ was
first realized by Murthy and Shankar \cite{qd-us2}. Not only did
this mean that the one loop flow of Murthy and Mathur was exact,
it meant the disorder-interaction problem of the chaotic dot could
be solved exactly in the large $g$ limit. It is the only known
case where the problem of disorder and
interactions\cite{intdisorder,belitz,weak-ferro} can be handled
exactly.

  From Eqn.
(\ref{rg-spinless2}) it can be seen that positive initial values
of $\tu_m$ (which are equal to initial values of $u_m$ inherited
from the bulk) are driven to the fixed point at $\tu_m=0$, as are
negative initial values as long as $u_m(t=0)\ge u_m^*=-1/\ln{2}$.
Thus, the Fermi liquid parameters are {\it irrelevant} for this
range of starting values. Recall that setting all $u_m=0$ for
$m\ne0$ results in the Universal Hamiltonian. Thus, the range
$u_m\ge u_m^*$ is the basin of attraction of the Universal
Hamiltonian.  On the other hand, for $u_m(t=0)\le u_m^*$ results
in a runaway flow towards large negative values of $u_m$,
signalling a transition to a phase not perturbatively connected to
the Universal Hamiltonian.

Since we have a large $N$ theory here (with $N=g$), the one-loop
flow and the new fixed point at strong coupling are parts of the
final theory. \footnote{However the exact location of the critical point cannot
be predicted, as pointed out to us by Professor Peet Brower. The
reason is that the Landau couplings $u_m$ are defined at a scale
$E_L$  much higher than $E_T$ (but much smaller than $E_F$) and
their flow till we come down to $E_T$, where our analysis begins,
is not within the regime we can control. In other words we can
locate $u^*$ in terms of what couplings we begin with, but these
are  the Landau parameters renormalized in a nonuniversal way as we come down from
 $E_L$ to $E_T$. }What is the nature of the  state for $u_m(t=0)\le
u_m^*$? The formalism and techniques needed to describe that are
beyond what was developed in these lectures, which has focused on
the RG. Suffice it to say that it is possible to write the
partition function in terms of a new collective field  $\sigmab$ (which depends on all the particles)
 and that the
action $S(\sigmab )$ has a factor $g$ in front of it, allowing us
to evaluate the integral by saddle point(in the limit  $g\to \infty$), to confidently predict
the strong coupling phase and many of its properties.
 Our expectations based on the large $g$
analysis have been amply confirmed by detailed numerical studies
\cite{gang4}. For now I will briefly describe the new phenomenon
in qualitative terms for readers not accustomed to these ideas and
give some references for those who are.

In the strong coupling region
$\sigmab$  acquires an expectation
value in the ground state. The dynamics of the fermions is
affected by this variable in many ways: quasi-particle widths
become  broad  very quickly above the Fermi energy, the second
difference $\Delta_2$ has occasionally very large values and can
even be negative, \footnote{How can the cost of adding one
particle be negative (after removing the charging energy)? The
answer is that adding a new particle sometimes lowers the energy
of the collective variable which has a life of its own. However,
if we turn a blind eye to it and attribute all the energy to the
single particle excitations, $\Delta_2$ can be negative.}, and the
system behaves like one with broken time-reversal symmetry if $m$
is odd.

Long ago Pomeranchuk \cite{pomeranchuk} found that if a Landau
function of a pure system exceeded a certain value, the fermi
surface underwent a shape transformation from a circle  to an
non-rotationally invariant form. Recently this transition has
received a lot of attention\cite{varma,oganeysan} The transition
in question is a disordered version of the same. Details are given
in Refs. (\cite{qd-us2}, \cite{gang4}).

Details aside, there is another very interesting point: even if
the coupling does not take us over to the strong-coupling phase,
we can see vestiges of the critical point $u_{m}^{*}$ and
associated critical phenomena. This is a general feature of many
{\em quantum critical points}\cite{critical-fan}, i.e., points
like $u_{m}^{*}$ where as a variable in a hamiltonian is varied,
the system enters a new phase (in contrast to transitions wherein
temperature $T$ is the control parameter).

Figure \ref{fan} shows what  happens in a generic situation. On
the $x$-axis a variable ($u_m$ in our case ) along which the
quantum phase transition occurs. Along $y$ is measured a new
variable, usually temperature $T$. Let us consider that case
first. If we move from right to left at some value of $T$, we will
first encounter physics of the weak-coupling phase determined by
the weak-coupling fixed point at the origin. Then we cross into
the {\em critical fan} (delineated by the $V$-shaped dotted
lines), where the physics controlled by the quantum critical
point. In other words we can tell there is a critical point on the
$x$ -axis without actually traversing it. As we move further to
the left, we reach the strongly-coupled symmetry-broken phase, with a non-zero
order parameter.

It can be shown that in our problem, $1/g$ plays the role of $T$.
One way to see is this that in any large $N$ theory $N$ stands in
front of the action when written in terms of the collective
variable. That is true in this case well for $g$. (Here $g$ also
enters at a subdominant level inside the action, which makes it
hard to predict the exact shape of the critical fan. The bottom
line is that we can see the critical point at finite $1/g$. In
addition one can also raise temperature or bias voltage to see the
critical fan.

Subsequent work has shown, in more familiar examples that Landau
interactions,  that the general picture depicted here is true in
the large $g$ limit: upon adding sufficiently strong interactions
 the Universal
Hamiltonian gives way to other descriptions with broken
symmetry\cite{brower,gm}. \footnote{The only result that is not
exact in the large $g$ limit is the critical value $u_m^*$ since
the input value of $u_m$ at $E_T$ is related in a non-universal
way to the Landau parameter. In other words, the Landau coupling
$u_m$ settles down to its fixed point value at an energy scale
that is much larger than $E_T$. We do not know how it flows as we
reach energies inside $E_T$ wherein our RMT assumptions are
valid.}

\begin{figure}
\centering
\includegraphics[height=4cm]{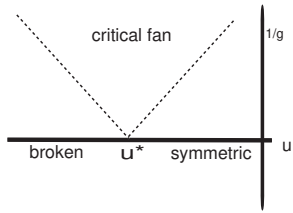}
%
%
\caption{The generic phase diagram for a second-order quantum
phase transition. The horizontal axis represents the coupling
constant which can be tuned to take one across the transition. The
vertical axis is usually the temperature in bulk quantum systems,
but is $1/g$ here, since in our system one of the roles played by
$g$ is that of the inverse temperature.}
\label{fan}       
\end{figure}

\section{Acknowledgement}
It is a pleasure to thank the organizers of this school especially
Professors Dieter Heiss, Nithaya Chetty and Hendrik Geyer for
their stupendous hospitality that made all of decide to revisit
South Africa as soon as possible.



\printindex
\end{document}